\documentclass[a4paper,11pt]{article}

\usepackage[ansinew]{inputenc}
\usepackage{graphicx} 
\usepackage[lmargin=1in, rmargin=1in,bottom=1in,top=1in]{geometry} 
\usepackage{amsmath}
\usepackage{amsthm} 
\usepackage{amssymb}  
\usepackage{url}
\usepackage[T1]{fontenc}
\usepackage{color}
\usepackage[rgb]{xcolor}
\usepackage{algorithm}
\usepackage{algorithmic}
\usepackage{psfrag}
\usepackage{tikz}
\def\eps{\varepsilon}
\def\p{\varphi}

\def\S{\mathcal{S}}
\def\U{\mathcal{U}}
\def\P{\mathcal{P}}
\def\C{\mathcal{C}}
\def\E{\mathcal{E}}
\def\V{\mathcal{V}}

\def\G{\mathcal{G}}
\def\F{\mathcal{F}}

\def\W{\mathcal{W}}
\def\Q{\mathcal{Q}}

\def\K{\mathcal{K}}

\def\s{\overline{s}}

\newtheorem{thm}{Theorem}

\newtheorem{conj}{Conjecture}

\begin{document}

\title { Policy Iteration is well suited to optimize PageRank }
\author{ 
			Romain Hollanders\thanks{Department of Mathematical Engineering, ICTEAM, UCLouvain, 4, avenue Lemaitre, B-1348 Louvain-la-Neuve, Belgium. This work was supported by the ARC grant 'Large Graphs and Networks' from the French Community of Belgium and by the IAP network 'Dysco' funded by the office of the Prime Minister of Belgium. The scientific responsablity rests with the authors.} \thanks{Corresponding author, \texttt{romain.hollanders@uclouvain.be} } \\
			 Jean-Charles Delvenne$^*$\thanks{\texttt{jean-charles.delvenne@uclouvain.be}}\\
			   Rapha\"el Jungers$^*$ \thanks{\texttt{raphael.jungers@uclouvain.be}}
	}
%\date{\emph{June 2011}}
\date{July 2011}
\maketitle

\setlength{\parindent}{0pt}
\setlength{\parskip}{1ex plus 0.5ex minus 0.2ex}

\vspace{-1cm}

\begin{abstract}

\noindent The question of knowing whether the policy Iteration algorithm (PI) for solving Markov Decision Processes (MDPs) has exponential or (strongly) polynomial complexity has attracted much attention in the last 50 years.  Recently, Fearnley proposed an example on which PI needs an exponential number of iterations to converge.
Though, it has been observed that Fearnley's example leaves open the possibility that PI behaves well in many particular cases, such as in problems that involve a fixed discount factor, or that are restricted to deterministic actions. In this paper, we analyze a large class of MDPs and we argue that PI is efficient in that case.  The problems in this class are obtained when optimizing the PageRank of a particular node in the Markov chain.  They are motivated by several practical applications.

We show that adding natural constraints to this PageRank Optimization problem (PRO) makes it equivalent to the problem of optimizing the length of a stochastic path, which is a widely studied family of MDPs.
Finally, we conjecture that PI runs in a polynomial number of iterations when applied to PRO. We give numerical arguments as well as the proof of our conjecture in a number of particular cases of practical importance. 

%\noindent We see this work as a step towards deeper insight on the link that can be made between the properties of an MDP instance and the resulting efficiency of PI.

\end{abstract}

\section*{Introduction}

In search engines, it is critical to be able to compare webpages according to their relative importance, with as few as possible computational resources. This is done by computing the \textit{PageRank} of every webpage from the web \cite{brin1998}~: pages with higher PageRank will then appear higher in the list of results. To compute this PageRank, the first step is to model the web as a digraph in which the webpages are represented by nodes and the links between them are represented by directed edges. Then, the PageRank of a node is defined as the average portion of time spent in that node during an infinite and uniform random walk on the graph. This random walk can be seen as the infinite process of a random surfer that, from its current page, picks up any available outgoing link with uniform probability and jumps to the page pointed by that link. 
%Formally, we define that process as a Markov chain where the transition between nodes is driven by a uniform transition matrix $P$ such that the probability of going from node $i$ to node $j$ $P_{ij} = 1/\mathrm{deg}(i)$ if there is a link from node $i$ to node $j$ and $P_{ij} = 0$ otherwise, where $\mathrm{deg}(i)$ is the out-degree of node $i$. In this setting, we assume that each jump of the random surfer has unit-cost.

The utility of PageRank is not limited to search engines and it has been proposed in several other applications such as financial market, spam detection, web-crawling, semantic networks, and many others. It can also be used in any application that requires ranking nodes in order of relative importance. See \cite{Berkhin2005} for a survey on PageRank and its applications. The introduction of the concept of PageRank has also generated a large number of questions and challenges. Among these, the problem of optimizing the PageRank of webpages raises increasing interest, as evidenced by the growing literature on the subject \cite{Avrachenkov2004,mathieu2006,DeKerchove2008,Tempo2009,Jungers2009bis,fercoq2010}. It is also of great practical interest and well-studied in the engineering community where good practice methods are developed to ensure a high PageRank \cite{chaffey2009}. PageRank Optimization (PRO) is also the focus of this paper. Here, we study how to maximize (or minimize) the PageRank of some target node when control is granted on some subset of edges, meaning that some edges (called the \textit{free edges}) may be chosen to be activated or deactivated. A typical example of PRO is the so-called \textit{webmaster problem} in which a webmaster tries to maximize the PageRank of one of his webpage by determining which links under his control (i.e. on his website, or on an allied website for instance) he should activate and which links he should not \cite{Avrachenkov2004,DeKerchove2008}. Furthermore, the same tools may be used to find how much the PageRank of some nodes can vary when the presence or absence of some links, called \textit{fragile links}, is uncertain (e.g. because a link is broken, the server is down or because of traffic problems) \cite{Tempo2009}.

The main difficulty to solve PRO is to deal with the exponential number of possible free edges configurations~: since each edge has two possible states - on or off - the number of possible configurations is $2^f$, where $f$ is the number of free edges. To escape this difficulty, Ishii and Tempo first proposed an approximate algorithm that would find an interval containing the minimum and maximum PageRank of the target node \cite{Tempo2009}. One year later, Cs\'aji et al. proposed a way of formulating the problem as a \textit{Stochastic Shortest Path} problem (SSP) - which is a subclass of \textit{Markov Decision Processes} (MDPs) - thereby showing that an exact solution of the problem could be found in weakly polynomial time using linear programming \cite{Jungers2009bis}. (For some refinements to PRO, see also \cite{fercoq2010}. For more on SSPs and MDPs see e.g. \cite{Puterman1994}, \cite{Bertsekas1991} and \cite{Bertsekas2007}.) Yet in practice, MDPs (and thus also SSPs) are solved much more efficiently using algorithms adapted to their special structure. Among these algorithms, \textit{Policy Iteration} (PI) \cite{Howard1960} performs amazingly well and is guaranteed to converge to the optimal solution in a finite number of iterations. However, even though PI usually converges in few iteration, theoretical upper and lower bounds on its complexity are exponential in many cases. The main goal of this paper is to show that the existing exponential lower bounds, as they are, should not apply to PRO. Instead, we believe that polynomial upper bounds exist in that case.

%If applied to PRO, one could describe the process of PI as follows : first we choose an arbitrary configuration of free edges. Then the algorithm iteratively repeats two steps : (1) an \textit{evaluation step} in which the quality of the configuration is computed (i.e. the corresponding PageRank of the target node) and (2) an \textit{improvement step} in which the evaluation is used to compute a new strictly improving configuration. By repeating these two steps, one eventually reaches the optimal configuration. 
There is a significant research effort for understanding the complexity of PI. Let us quickly review existing complexity results. For general MDPs, the best upper bound - $O(2^m/m)$ - is due to Mansour and Singh \cite{Mansour1999}, where $m$ designates the number of choices to be made in the problem, while the largest lower bound is also exponential and has recently been found by Fearnley through a carefully built example \cite{Fearnley2010}. This was a breakthrough after 50 years of research on the question of the complexity of PI. The story is different for discounted MDPs (i.e. a class of MDPs in which the impact of future costs are progressively reduced by some discount factor) for which a first strongly polynomial upper bound has recently been found by Ye \cite{Ye2010} and then further improved by Hansen et al. \cite{hansen2010B},
% to $O(\frac{n}{1-\gamma} \log \left( \frac{m}{1-\gamma} \right))$, where $n$ is the number of nodes
yet only for fixed discount factors. Though, even if upper and lower bounds seem to meet in both cases, the story does not end here. Indeed, Fearnley's example is impossible to adapt to some other important particular cases of MDPs. This is for example the case for Deterministic MDPs (DMDPs) for which the best lower bound currently known is quadratic and has been found by Hansen and Zwick \cite{hansen2010}. Besides, strongly polynomial time algorithms exist to solve that problem \cite{madani2010} (including PI when fixed discount is included to the problem \cite{Ye2010}), which hints that DMDPs might be easier to solve than general MDPs. 

In this work, we argue that PRO might be another particular case for which a polynomial reduction of Fearnley's example is not possible. We first show how a natural generalization of PRO makes it equivalent to general SSPs and vice versa, giving a new point of view on SSPs (and MDPs). Then, we identify the exclusive nature of the choice of actions in an SSP as the main constraint that makes it different and most probably harder to solve than PRO. Based on extensive numerical computations, we then conjecture that PI converges in a polynomial number of iterations in the case of PRO. We also give a number of particular cases in which we show that PI converges in polynomial time. 
In this work, we try to make a step towards deeper insight on the existing link between the properties of an MDP instance and the resulting efficiency of PI.
%We see this work as a new step towards a classification of MDPs between the ones that can be solved in (strongly) polynomial time using PI, and the ones that cannot.

The paper is organized as follows. In Section 1, we give formal definitions for SSP and PRO and we generalize PRO in a natural way. In Section 2, we show how that generalization of PRO can be transformed into an SSP and vice versa. In Section 3, we conjecture that PI should perform well when applied to PRO, arguing with numerical evidence. Then, in Section 4, we give a number a particular cases of PRO for which PI behaves well.

\section{Definitions} \label{sec2}

In this section, we give a formal definition for Stochastic Shortest Path and for PageRank Optimization problems. We also formulate a natural generalization of the latter.

\textbf{Stochastic Shortest Path. } 
An instance of the \textit{Stochastic Shortest Path} problem (SSP) is a tuple $(\S, \U, \P, \C)$ where $\S$ is the finite set of \textit{states}, $U$ is the finite set of all \textit{actions} and $\U_s \subseteq \U$ is the set of actions available in state $s \in \S$ (there is at least one action for each state), and $\P_{s,s'}^{u}$ and $\C_{s,s'}^{u}$ are respectively the \textit{transition probability} of going from state $s \in \S$ to state $s' \in \S$ when choosing action $u \in \U_s$ and the (real-valued) \textit{cost} incurred by this displacement \cite{Bertsekas1991}. 
%So, in state $s \in \S$, choosing an action $u \in \U_s$ will induce a transition to some other states $s' \in \S$, each of which being chosen with some probability defined by $\P$ and inducing some costs defined by $\C$. 
We also ask for the transition probabilities to be non-negative and to sum to one, namely $\sum_{s' \in \S} \P_{s,s'}^u = 1$, for all starting state $s \in \S$ and action $u \in \U_s$. An action is said \textit{probabilistic}  if it includes randomization between several arrival states, whereas it is said \textit{deterministic} otherwise. 

In SSP, we consider the random process of an \textit{agent} that starts at some starting state $s_0$ and then jumps to a new available state at each time step (according to the action taken in its current state and the associated transition probabilities). The main feature of an SSP, compared to general MDPs, is that we assume the existence of an absorbing cost-free state $\tau$ (also called \textit{target} state) that is required to be reachable with a non-zero probability path by every other state, whichever actions are chosen. In this context, the goal of the \textit{controller} of the process is to choose the right action in each state in order to minimize the expected sum of costs incurred by the agent before reaching the target state, whatever the starting state. The choice of a unique action to take in each state is called a \textit{policy} (or strategy) $\mu : \S \rightarrow \U$. The chosen policy is \textit{proper} if the agent eventually reaches $\tau$ for any starting state. It is \textit{improper} otherwise. 
%Furthermore, a policy $\mu$ is \textit{positional} if it is deterministic and history-independent.
A policy is \textit{optimal} iff it is better at minimizing the controller's goal than any other policy, for any starting state.
%To compare policies in terms of their ability to serve the controller's goal, we define the \textit{value} $V^{\mu} : \S \rightarrow \R$ of a policy $\mu$ as the sum of costs incurred by the agent when following that policy for each starting state $s \in \S$. In SSP, these values are \textit{well defined} (i.e. finite) provided the considered policies are proper. We say that a policy $\mu$ \textit{dominates} $\mu'$ iff $V^{\mu}(s) \leq V^{\mu'}(s)$ for all states $s \in \S$. A policy $\mu$ is \textit{optimal} iff it dominates any policy $\mu'$. 
One fundamental result about MDPs that can be adapted to SSPs guarantees that there always exists an optimal (not necessarily unique) proper policy, provided that there exists at least one proper policy \cite{Puterman1994,Bertsekas2007}. Note that it is always possible to formulate an SSP problem as a linear program whose size is polynomial in the number of states and the maximum number of actions per state of the SSP instance.

\textbf{PageRank Optimization. } 
To define a \textit{PageRank Optimization} problem (PRO), we first define its \textit{support graph} $\G = (\V, \E)$, where $\V = \V' \cup v$ is the set of nodes, $\E \subseteq \V \times \V$ is the set of directed edges of the graph and $v$ is the target node for which we want to maximize (or minimize) the PageRank. For that task, control is granted on some subset $\F \subseteq \E$ of edges (called the set of \textit{free edges}) in which we may choose to activate or deactivate any edge, whereas the edges in $\E \backslash \F$ are fixed and cannot be removed. The goal in PRO is to choose the right subset of free edges that we activate so that the PageRank of node $v$ is maximal (or minimal). Here we focus on the maximization problem but a straightforward modification of the approach can be used to deal with the minimization problem as well.

%So, the goal in a PRO problem is to find the best way of partitioning $\F$ into a set of activated edges $\F^+$ and a set of deactivated edges $\F^- = \F \backslash \F^+$ such that the PageRank of $v$ in the subgraph $(\V, \E \backslash \F^-)$ is maximal (or minimal). Here we focus on the maximization problem but a straightforward modification of the approach can be used to deal with the minimization problem as well.

PRO can be formulated as an SSP in polynomial time \cite{Jungers2009bis} as follows. First observe that maximizing the PageRank of $v$ (i.e. its frequency of visit by the random surfer) is equivalent to minimizing the average time between two visits of $v$. Let us split $v$ into a starting node $v_s$ and a target node $v_t$ such that $v_s$ has all outgoing links of $v$ and $v_t$ has all its ingoing links, plus a zero-cost self-loop. Maximizing the PageRank of $v$ is then equivalent to minimizing the average distance from $v_s$ to $v_t$. Observe that $v_t$ is now an absorbing node. In this setting, an action is the choice between activation or deactivation for a given free edge and a policy is a subset of activated free edges. %So if there are $n$ nodes and $f$ free edges
%Furthermore, the activation or deactivation of a free edge can be seen as an action and the choice of a subset of activated free edges can be seen as a policy. 
Therefore, PRO can be seen as a particular case of SSP where $\overline{\V} = \V' \cup \{ v_s, v_t \}$ is the set of states and $\F$ is the set of $2f$ actions\footnote{For precision, taking an action in a state in which $k$ outgoing edges are free should be seen as choosing the subset of these $k$ edges that should be activated. See Section \ref{sec3} for a construction to deal with the size of these subsets in polynomial time.}, where $f$ is the number of free edges. This SSP instance has a polynomial number of states and actions. Uniform transition probabilities and unit costs are assumed here (except for the target node $v_t$ which is cost-free). 
%Given a policy $\mu$ (i.e. a configuration of free edges), we obtain a subgraph $\G^{\mu} = (\V' \cup \{ v_s, v_t \}, \E^{\mu})$, where $\E^{\mu}$ is the set of activated edges corresponding to policy $\mu$.

If we do not assume that the support graph $\G$ is strongly connected, extra care must be taken. Indeed, in that case, there may be nodes (or connected components) that do not have any outgoing edge. To deal with such \textit{dangling} nodes (or components), many techniques exist \cite{Berkhin2005}, such as connecting them to every other node. We may choose any of the existing solution and assume that we have already dealt with dangling nodes. Another case for which we must be careful is when all outgoing edges from a node are free. Indeed, such a node would become a dangling node is every free edge was deactivated. However, in that case, Csáji et al. have shown that the optimal policy would always activate exactly one of these free edges, i.e. the one that points towards the node that is closest to the target node \cite{Jungers2009bis}. Furthermore, an algorithm like PI would only consider policies in which exactly one of these free edges are active. Therefore, in this case, we may always consider that there are as many actions as free edges, where each action consists in activating exactly one of the free edges. As a consequence of the above, we may always consider that all nodes are able to reach the target node with positive probability, whatever the chosen policy (so that every policy is proper).

\textbf{Generalized PageRank Optimization. } 
A natural way of extending PRO is to allow arbitrary transition probabilities and costs. We call such a relaxation a \textit{Generalized PageRank Optimization} problem (GPRO), which we now formally define. A convenient way to deal with arbitrary transition probabilities in the context of GPRO, where the out-degree of the nodes may vary, is to assign a weight to each possible transition and to compute the transition probabilities in proportion to these weights. More precisely, we define the weight set $\W$ such that $\W_{i,j} > 0$ if $(i, j) \in \E$ and $\W_{i,j} = 0$ otherwise. Given a policy $\mu$ (i.e. a configuration of free edges), we define the corresponding weight set $\W^{\mu}$ as
\begin{equation*}
	\W_{i,j}^{\mu} = 
	\begin{cases}
		\W_{i,j} \hspace{.8cm}   \text{if } (i, j) \in \E^{\mu}, \\
		0        \hspace{1.3cm} \text{otherwise},
	\end{cases}
\end{equation*}
where $\E^{\mu}$ is the set of activated edges when policy $\mu$ is chosen. Transition probabilities $\Q^{\mu}$ are then defined according to the weights by :
\begin{equation*}
	\Q_{i,j}^{\mu} = \frac{\W_{i,j}^{\mu}}{\sum_{k \in \overline{\V}} \W_{i,k}^{\mu}}.
\end{equation*}
In a node $i$, weights enable to distribute the probabilities among the activated edges that leave $i$, and this in proportion to their mutual importance. Note that for $\Q$ to be well defined, there must always exist at least one edge with positive weight going out of any starting node $i$, for any policy $\mu$. For any nodes $i, j$, a cost matrix $\K$ is also defined such that $\K_{i,j}$ is the cost of going from $i$ to $j$. 

Finally, in the GPRO framework, we also allow some exclusivity constraints in the following case~: in a node in which there are only two free edges and no fixed edges going out, we may assume that exactly one of these edges must be activated while the other must be deactivated. We will see that these exclusivity constraints will enable us to make the link with SSP. We believe that such constraints are the key difference with PRO that makes the latter problem easier to solve.
%In an abstract way, this can be seen as if one of the two edges was free with high weight and the other was fixed with negligible weight, so that activating the free edge means ``taking this edge'' while deactivating the free edge means ``taking the fixed edge''. So, assuming exclusivity constraints is equivalent to assuming that weights may be arbitrarily small and it is still a natural way of extending PRO.

Putting everything together, we define a GPRO instance by the tuple $(\overline{\V}, \F, \W, \K)$. The original edge set $\E$ can be obtained from $\W$. Let us now make some comments about the introduced concepts.
%, while the nodes $v_s$ and $v_t$ can for instance be chosen as the first and last node respectively WLOG. 

\textbf{Remarks.}

\vspace{-.4cm}
\begin{enumerate}
	\item A PRO problem can be formulated as a GPRO problem in which $\W_{i,j} = 1$ if $(i, j) \in \E$ and $\W_{i,j} = 0$ otherwise, and in which $\K_{i,j} = 1$ for all $(i, j) \in \E$ (except when $i = v_t$). 
	\item Exclusivity constraints can be modeled using small weights. Indeed, in a node where there is a free and a fixed edge, if the free edge has a weight significantly higher than the fixed edge, it means that this edge will be chosen with high probability if activated, while the fixed edge will always be chosen with probability one if the free edge is deactivated : so depending on the activation state of the free edge, one edge or the other will be chosen, which imitates the exclusive behavior of SSP actions, as illustrated in figure \ref{SSPactionGPRO}. However at that point, we have not been able to adapt such a model to any instance with the guarantee that the optimal solution would not change, at least not with weights that have polynomial value.
	
	\begin{center}
	\vspace{-.3cm}
	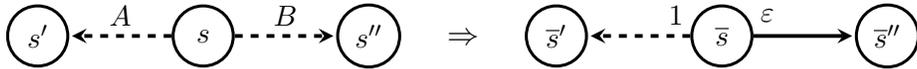
\begin{figure}[!ht]
		\begin{center}
		\begin{tabular}{ccc}
		
			\begin{tikzpicture}[scale=1.1,>=stealth,->,shorten >=1pt,looseness=1,auto]
				\begin{scope}[every node/.style={font=\small\itshape},line width=0.5mm]
					\tikzstyle{every node}=[shape=circle,line width=0.4mm,minimum size=.8cm];
					\path (0,0)  node[draw] (i)  {$s$};
					\path (-2,0) node[draw] (i1) {$s'$};
					\path (2,0)  node[draw] (i2) {$s''$};
					\draw[->,dashed] (i) -- (i1) node [midway,above,draw=none,yshift=-.15cm] {$A$};
					\draw[->,dashed] (i) -- (i2) node [midway,above,draw=none,yshift=-.15cm] {$B$};
				\end{scope}
			\end{tikzpicture}
			&
			\begin{tikzpicture}[scale=1.1,>=stealth,->,shorten >=1pt,looseness=.8,auto]
				\begin{scope}[every node/.style={font=\small\itshape},line width=0.5mm]
					\tikzstyle{every node}=[shape=circle,line width=0.4mm,minimum size=.8cm];
					\path (0,0) node (a) {\large{$\Rightarrow$}};
%					\draw[color=white] (0,-.5)--(0,-2.35);
				\end{scope}
			\end{tikzpicture}
			&
			\begin{tikzpicture}[scale=1.1,>=stealth,->,shorten >=1pt,looseness=1,auto]
				\begin{scope}[every node/.style={font=\small\itshape},line width=0.5mm]
					\tikzstyle{every node}=[shape=circle,line width=0.4mm,minimum size=.8cm];
					\path (0,0)  node[draw] (i)  {$\s$};
					\path (-2,0) node[draw] (i1) {$\s'$};
					\path (2,0)  node[draw] (i2) {$\s''$};
					\draw[->,dashed] (i) -- (i1) node [near start,above,draw=none,xshift=.15cm,yshift=-.15cm] {$1$};
					\draw[->] (i) -- (i2) node [near start,above,draw=none,xshift=-.15cm,yshift=-.15cm] {$\eps$};
				\end{scope}
			\end{tikzpicture}
			
		\end{tabular}
		\caption{\footnotesize Left : exclusive actions in the GPRO framework. The controller is asked to choose either $A$ or $B$. Right : the equivalent action modeled with weights in the GPRO framework. The controller is asked to activate or deactivate the free (dashed) edge. Here, $\eps$ represents a ``small enough'' weight. } \label{SSPactionGPRO}
		\end{center}
	\end{figure}
	\vspace{-.3cm}
\end{center}

	\item In SSP and GPRO, exclusivity constraints concern edges that leave the same node. Cs\'aji et al. have shown in \cite{Jungers2009bis} that if one adds exclusivity constraints between free edges that leave different nodes, PRO becomes NP-hard to solve. This fact is another clue towards the fact that these constraints do make a difference in the efficient solvability of these problems.
	\item Solving a GPRO problem can be seen as the search of the best subgraph of a given graph (the support graph) such that some edges cannot be removed, and such that it satisfies additional exclusivity constraints.
	%graph that is a subgraph of the support graph and a supergraph of the same support graph in which we deactivate every free edge. Of course, the searched graph must satisfy the exclusivity constraints, if any.
	\item Because of the context of PRO, we focused here on an SSP-like criterion in which an absorbing state has to be reached as quickly as possible. However, the GPRO formulation could have been adapted to match any MDPs' classical optimization criteria, like the average-cost and the discounted-cost criteria for instance.
\end{enumerate}

\section{Comparison between PRO and SSP} \label{sec3}

In this section, we show that from any instance of SSP, we can build a GPRO instance that has the same optimal solution, and vice versa. 

\begin{thm}

Given any instance of an SSP problem with $n$ states and a total of $m$ available actions, it can be reduced in polynomial time to a GPRO problem with $O(m)$ nodes and $O(m)$ free edges that has the same optimal solution. Similarly, given any instance of a GPRO problem with $n$ nodes and $f$ free edges, it can be transformed in polynomial time into an SSP problem with $O(n)$ single-action states and $f$ 2-actions states that has the same optimal solution. 

\begin{proof} 

We first show how to reduce an SSP to a GPRO and then a GRPO to an SSP.
%, we observe how an exclusive choice between two actions may be represented in the GPRO framework.

\textbf{\textsc{Going from SSP to GPRO. }} Let us create a GPRO instance with a set of nodes $\overline{\V}$ that corresponds to the set of states $\S$ of the SSP. Then, we first claim that any SSP can be expressed as another SSP problem in which there are at most two actions per state, without changing the optimal solution. Then we show how probabilistic 2-actions states can be split into one deterministic 2-actions state and two single-action states. We conclude by showing how single-action states and deterministic 2-actions states can be reduced in the GPRO framework. 

\textit{\textbf{Claim 1 :}} Given any state $s$ of an SSP instance with $k \geq 2$ available actions, $s$ can be split into $(k-1)$ 2-actions states without changing the optimal solution of the original SSP. We show this by induction on $k$. The base case for $k = 2$ is trivial. Then, if it is true for $k-1$, it is still true for $k$. Indeed, let us split $s$ into two states $s'$ and $s''$ and suppose that $s''$ has $(k-1)$ actions that correspond to the last $(k-1)$ actions of $s$ while $s'$ has two actions : one that corresponds to the first action of $s$ and one that goes to state $s''$ deterministically with probability $1$ and cost $0$. Actions that were previously pointing towards $s$ are now pointing towards $s'$. Hence state $s'$ corresponds to state $s$ but it has a restricted decision to take : either the first action of $s$ or some of the other actions. The optimal action to take in $s$ does not change in this construction since if the first action of $s$ was optimal, it will also be taken in $s'$ and if not, it means that some of the other actions of $s$ would be preferable so the decision is postponed by choosing the second action of $s'$ that goes to $s''$. Since $s''$ has $(k-1)$ available actions, it can be split into $(k-2)$ 2-actions states without changing the optimal solution by induction hypothesis, which makes a total of $(k-1)$ 2-actions states.

\textit{\textbf{Claim 2 :}} A probabilistic 2-actions state $s$ of an SSP instance can be split into one deterministic 2-actions state $u$ and two (probabilistic) single-action states $u'$ and $u''$. In $u$, the choice of one of the two available actions is done, allowing the process to move deterministically to either $u'$ or $u''$ with probability $1$ and cost $0$. Now the only available action in $u'$ (resp. $u''$) performs the randomization relative to the first (resp. second) action of $s$. 

Using Claims 1 and 2, we transform the original SSP problem with $n$ states and a total of $m$ actions into an equivalent SSP problem with $O(m)$ states, all of them being either deterministic 2-actions states or probabilistic single-action states. Indeed, we first create an equivalent SSP with only 2-actions states using Claim 1 and then we transform every probabilistic 2-actions state into one deterministic 2-actions states and two probabilistic single-action states using Claim 2. These transformations can be done in polynomial time and the resulting SSP problem is equivalent to the first SSP in the sense that it still has the same optimal solution. We now give the tools to transform this new SSP problem into a GPRO.

\textit{\textbf{Claim 3 :}} In an SSP instance, a deterministic action in state $s$ in which one must choose either action $A$ or action $B$ can be reproduced using exclusivity constraint in the GPRO setting. We saw in Section \ref{sec2} that this is done by giving two free edges to the node $\s$ that corresponds to state $s$, and assuming that these edges are linked with an exclusivity constraint. So, 2-actions states in the SSP setting can be modeled using two free edges in the GPRO setting.

\textit{\textbf{Claim 4 :}} A single-action state $s$ that randomizes between a set of states $\S'$ can be reduced to the GPRO framework by adding an edge from the node $\s$ corresponding to $s$ to every node $\s'$ that correspond to the states in $\S'$. To every such edge $(\s, \s')$, we give a weight $\W_{\s,\s'} = \P_{s,s'}$ and a cost $\K_{\s,\s'} = \C_{s,s'}$. It is not hard to see that the obtained randomization effect in GPRO is equivalent to that of the SSP (same transition probabilities and same costs). Hence, single-action states can be modeled without any free edge.

%\textbf{Claim 3 :} A deterministic 2-actions state $s$ in SSP can be reduced to a node $\overline{s}$ with two mutually exclusive free edges. 
%If in state $s$, action $A$ in state $s$ goes with probability $1$ to state $s'$ and action $B$ goes with probability $1$ to state $s''$, it is quite obvious that one free edge in node $\overline{s}$ will go to node $\overline{s}'$ corresponding to state $s'$ with cost $\C_{s,s'}$ and weight 1 and that the other free edge will go to node $\overline{s}''$ corresponding to state $s''$ with cost $\C_{s,s''}$ and weight 1. Here, the introduction of an exclusivity constraint in GPRO copies exactly the idea of choosing exactly one action among two available in the SSP setting.
%Notice that these exclusivity constraints from GPRO can be handled in an appropriate algorithm in the same way actions are handled when tackling an SSP problem.

We may now reduce the new SSP problem obtained from Claims 1 and 2 into a GPRO, using Claims 3 and 4. Indeed, Claim 3 tells us how to reduce deterministic 2-actions states in a GPRO setting, whereas Claim 4 gives us the argument to reduce probabilistic single-action states, both in polynomial time. The resulting GPRO problem has $O(m)$ nodes (as many as the number of states of the transformed SSP) and $O(m)$ free edges ($2$ free edges for every $2$-actions state). In the resulting GPRO, all the nodes correspond to some state from the transformed SSP and they all have the same probability distribution and transition costs, so the optimal solution of both problems is identical. 
%Furthermore, each of the four steps of the construction can be performed in polynomial time.
 
%Putting everything together, from an SSP with $n$ states and a total of $m$ available actions, we first create another equivalent SSP with $O(m)$ states, all with at most 2 actions using claim 1. Then every probabilistic 2-action state is transformed into a deterministic 2-action state and two single-action states using claim 2. This transformed SSP still conserves the same optimal solution as the original SSP. Furthermore, it has now only single-action states and deterministic 2-actions states. Claims 3 and 4 give us the necessary tools to reduce the SSP to a GPRO with the same number of nodes and $O(m)$ free edges ($2$ free edges for every $2$-actions state). In the resulting GPRO, all the nodes correspond to some state from the transformed SSP and they all have the same probability distribution and transition costs, so the optimal solution of both problems is identical. Furthermore, each of the four steps of the construction can be performed in polynomial time.

\textbf{\textsc{Going from GPRO to SSP. }} A GPRO is already an instance of SSP with however a small distinction concerning the way the action set is described~: in GPRO, actions are taken in the free edges while in SSP, actions are taken in the states. However, we may assume that the actions in GPRO are also taken in the states but then, extra care must be taken. Indeed, if several free edges go out from one single node (say $k$ free edges), then every possible configuration of these free edges (so $2^k$ configurations) must be considered as an available action in that node and therefore, the number of actions per node can grow exponentially (in the worst case, we may end up with one node that has $2^f$ available actions, where $f$ is the number of free edges).

Before going further, we must consider the case of a node in which all outgoing edges are free. It that case, we saw in section \ref{sec2} that we may consider as many actions as there are free edges such that each action corresponds to a situation in which exactly one free edge is activated. Therefore, such nodes with $k$ outgoing free edges may be transformed into a state with $k$ actions.

Now let us consider a node $i$ with $k > 1$ outgoing free edges in addition to some fixed edges. We show that such nodes can be transformed into a substructure in which every node has at most two outgoing free edges. The main idea of the construction, illustrated at figure \ref{SingleFEperNode}, is to create a new artificial node for every outgoing free edge, which is designed to act exactly as the original free edge. In node $i$ the choice of any edge is taken with respect to their weight but independently from the activation state of the free edges. If a free edge is chosen, the process jumps to the corresponding auxiliary node with cost 0. If the edge was activated, the path that leaves the structure is taken with probability 1 and the cost that corresponds to the original edge while if it is not, the process returns to node $i$ with probability 1 and cost 0. This procedure is then repeated until an activated edge is chosen. Thus, since there is always at least one fixed outgoing edge, we are always able to leave the structure. Observe that the auxiliary nodes exactly match the nodes with exclusive constraints described in the definition of GPRO and they can thus be transformed into states with two deterministic actions in the SSP setting. 

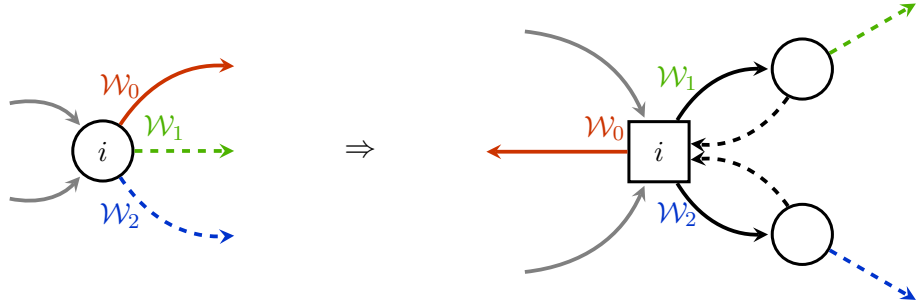
\begin{figure}[!ht]
	\vspace{-.7cm}
	\begin{center}
		\begin{tabular}{ccc}
		
			\begin{tikzpicture}[scale=1.1,>=stealth,->,shorten >=1pt,looseness=1,auto]
				\begin{scope}[every node/.style={font=\small\itshape},line width=0.5mm]
					\tikzstyle{every node}=[shape=circle,line width=0.4mm,minimum size=.8cm];
					\path (0,0)      node[draw]      (i)  {$i$};
					\path (2,1)      node[draw=none] (i1) {};
					\path (2,0)      node[draw=none] (i2) {};
					\path (2,-1)     node[draw=none] (i3) {};
					\path (-1.5,.5)  node[draw=none] (j1) {};
					\path (-1.5,-.5) node[draw=none] (j2) {};
					\draw[->,color=white!50!black] (j1) to [bend left] (i);
					\draw[->,color=white!50!black] (j2) to [bend right]  (i);
					\draw[->,color=green!20!red]	(i) to [bend left] (i1) node [at start,above,draw=none] {$\W_0$};
					\draw[->,dashed,color=red!30!green] (i) -- (i2) node [at start,above,draw=none,xshift=.4cm,yshift=-.2cm] {$\W_1$};
					\draw[->,dashed,color=green!20!blue] (i) to [bend right] (i3) node [at start,below,draw=none] {$\W_2$};
					\draw[color=white] (0,-1)--(0,-2.35);
				\end{scope}
			\end{tikzpicture}
			&
			\begin{tikzpicture}[scale=1.1,>=stealth,->,shorten >=1pt,looseness=.8,auto]
				\begin{scope}[every node/.style={font=\small\itshape},line width=0.5mm]
					\tikzstyle{every node}=[shape=circle,line width=0.4mm,minimum size=.8cm];
					\path (0,0) node (a) {\large{$\Rightarrow$}};
					\draw[color=white] (0,-.5)--(0,-2.35);
				\end{scope}
			\end{tikzpicture}
			&
			\begin{tikzpicture}[scale=1.1,>=stealth,->,shorten >=1pt,looseness=1,auto]
				\begin{scope}[every node/.style={font=\small\itshape},line width=0.5mm]
					\tikzstyle{every node}=[shape=circle,line width=0.4mm,minimum size=.8cm];
					\path (0,0)     node[draw,shape=rectangle] (i) {$i$};
					\path ( 30:2)   node[draw]      (i1) {};
					\path (-30:2)   node[draw]      (i2) {};
					\path (180:2.5) node[draw=none] (j0) {};
					\path ( 30:4)   node[draw=none] (j1) {};
					\path (-30:4)   node[draw=none] (j2) {};
					\path (-2, 1.5) node[draw=none] (j4) {};
					\path (-2,-1.5) node[draw=none] (j5) {};
					\draw[->,color=white!50!black] (j4) to [bend left] (i);
					\draw[->,color=white!50!black] (j5) to [bend right]  (i);
					\draw[->]	(i) to [bend left] (i1) node [at start,above,draw=none,color=red!30!green,yshift=.05cm] {$\W_1$};
					\draw[->]	(i) to [bend right] (i2) node [at start,below,draw=none,color=green!20!blue,yshift=.1cm] {$\W_2$};
					\draw[->,color=green!20!red] (i) -- (j0) node [at start,above,draw=none,xshift=-.3cm,yshift=-.2cm] {$\W_0$};
					\draw[->,dashed,color=red!30!green] (i1) -- (j1);
					\draw[->,dashed,color=green!20!blue] (i2) -- (j2);
					\draw[->,dashed] (i1) to [bend left] (.35,.1);
					\draw[->,dashed] (i2) to [bend right] (.35,-.1);
				\end{scope}
			\end{tikzpicture}
			
		\end{tabular}
		\vspace{-.7cm}
		\caption{\footnotesize In the right figure, there are maximum two free edges (i.e. dashed edge) per node, even though the two substructures have the same dynamics. All the costs of the black edges are zero, while the costs of the colored edges are the same as the costs of the edges of corresponding color in the left figure. 
%		The idea of the transformation is that each edge is first chosen according to its weight, independently from the activation state of the free edges. When a free edge is chosen, it is taken with probability 1 if it is activated, while the process returns to node $i$ if it is not, repeating the procedure until an activated edge is chosen.
		} \label{SingleFEperNode}
	\end{center}
	\vspace{-.5cm}
\end{figure}

The whole process needs a polynomial number of transformations (add one node and two edges for some free edges).
%and the obtained SSP being the same problem as the original GPRO, both problems have the same optimal solution.
\vspace{-.7cm}

\end{proof}

\end{thm}

%This theorem, and in particular claim 3, gives us the necessary intuition about why an SSP instance cannot be cast into a simple PRO instance. Indeed, claim 3 tells us that to reduce deterministic actions in a PRO framework, we must allow the weight of a free edge to be arbitrarily bigger than the weights of the other edges leaving the same node, which seems to be difficult, and even likely impossible to reproduce using a simple uniform random walk. In fact, the presence of such a uniform random walk should make it impossible to simulate free edges with more than half of the total weights in a node, at least not without creating ``dead-ends'' in the graph. We believe that this is the key difference between SSPs and PROs that make the latter problems easier to solve using Policy Iteration, as we will see in the next sections. Also note that stretching the weights to the limit as we did in figure \ref{SSPactionGPRO} can also be interpreted as if the two outgoing edges in node $\s$ were free edges with an exclusive constraint on them imposing one or the other to be activated in an exclusive way. This means that actions in an SSP are intrinsically exclusive, which must be put in relation with a result from \cite{Jungers2009bis} that shows that solving PRO with any exclusive constraints on the free edges is NP-complete.

As a final remark for this section, observe that all the arguments we have been using are not specific to the SSP optimization criterion. Here, we focused on SSP because the GPRO formulation originally comes from an SSP-like problem. However, it is easy to generalize GPRO to make it also equivalent to any MDP, whatever the chosen optimization criterion. The same arguments that we used here may be used to make the requested link. As a consequence, an MDP can always be formulated as the search of the best subgraph in a support graph (with some constraints on the edges that are allowed to be removed). This may be useful to enrich the way MDPs are usually viewed and enhance the associated intuition.
%observe that even though all arguments have been used to link GPRO to SSP, they would also be valid if a comparison with another kind of MDP was requested. 

\section{Applying Policy Iteration to PRO}

An adaptation of Policy Iteration (PI) to PRO has been proposed by Cs\'aji et al. in \cite{Jungers2009bis}. When writing the algorithm, we represent a configuration of free edges by the set of activated free edges that we denote by policy $\mu$. We also define the first hitting time $\p_i^{\mu}$ of node $i$ under policy $\mu$ as the average time needed to reach the target node $v_t$ when starting the process at node $i$ and following policy $\mu$ afterwards. Of course, $\p_{v_t}^{\mu} = 0$. First hitting times can be computed in polynomial time by solving a linear system.

%\newpage

We call the resulting adaptation of PI~: \textit{PageRank Iteration} (PRI). The different steps are formalized in Algorithm \ref{algPI}.

\algsetup{indent=2em}
\newcommand{\PI}{\ensuremath{\mbox{\sc PageRank Iteration}}}
\begin{algorithm}[h!]
	\caption{\textsc{PageRank Iteration}}
 	\label{algPI}
	\begin{algorithmic}[1]
		\REQUIRE An arbitrary policy $\mu_0$, $k = 0$.
		\ENSURE The optimal policy $\mu^*$.
		\medskip
		\WHILE {$\mu_k \neq \mu_{k-1}$}
			\STATE Evaluation step : compute $\p^{\mu_k}$.
			\STATE Greedy Improvement step : $\mu_{k+1} = \{ (i,j) \in \F : \p_i^{\mu_k} \geq \p_j^{\mu_k} + 1 \}$.
			\STATE $k \leftarrow k+1$.
		\ENDWHILE
		\RETURN $\mu_k$.
	\end{algorithmic}
\end{algorithm}

To summarize the operating mode of PRI, we start with an arbitrary policy and then proceed iteratively. At each iteration we determine the set of free edges that are such that, if they were independently switched (i.e. switched \textit{on} if edge is \textit{off} and vice versa), the resulting policy would improve on the preceding one. Then, PRI being a greedy version of PI, we make all the improving switches simultaneously, assuming that this would be even better than single switches. This procedure improves the policy iteratively until no more improvements are possible, meaning that it has converged to the optimal policy. Observe that one does not have to transform the PRO problem in an SSP problem, as the modifications of the policies are handled implicitly directly in the PRO problem.

Since each iteration needs polynomial time to compute, the only condition for PRI to run in polynomial time is to run in a polynomial number of iterations. Unfortunately, determining bounds on the number of iterations of PRI is an open question~: the best known upper bound $O(2^f/f)$ is adapted from Mansour and Singh \cite{Mansour1999} whereas Fearnley's exponential lower bound seems unlikely to apply to PRI for the reasons exposed in the previous sections.

We formulate the following conjecture, based on extensive computations.
%After a significant number of simulations and many unfruitful attempts to construct PRO instances on which PRI would need many iterations to converge, we believe the following conjecture to be true.

\begin{conj} \label{conj1}
The number of iterations of PRI is polynomial in the number of free edges.
\end{conj}

%It seems that the intrinsic probabilistic nature of PRO actions (i.e. because of the predefined uniform random walk) constitutes the essential difference with MDPs, where the choice of an action is intrinsically exclusive. 

\begin{figure}[!ht]
	\centering
	{
		\includegraphics[width=15cm]{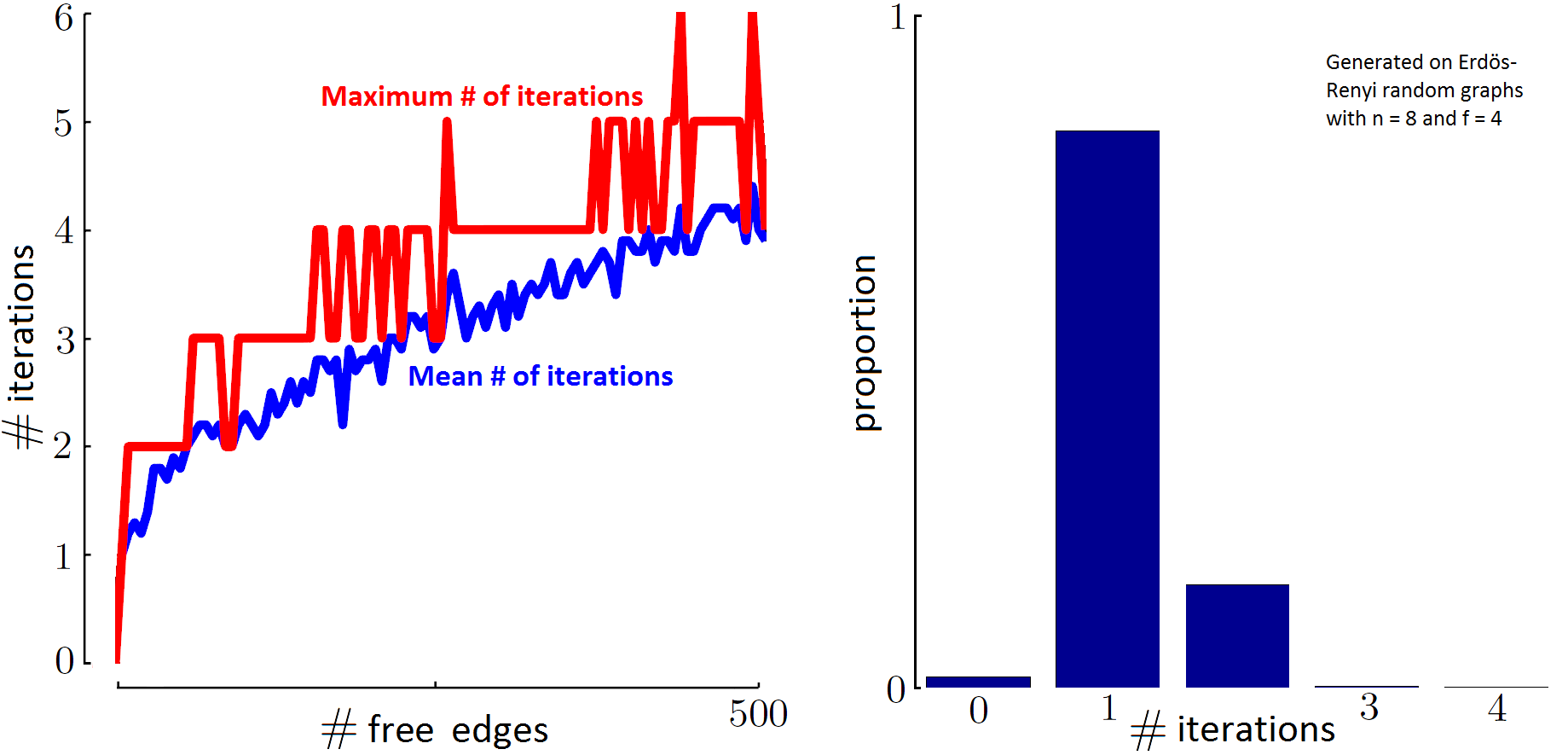}
	} 
	\caption{\footnotesize Left : the evolution of the number of iterations when the number of free edges grows. For each value of $f$, 5 tests have been performed (here on Power Law random graphs) and the average (in blue) and the maximum (in red) number of iterations have been recorded. Right : the distribution of the number of iterations of PRI after over 3 million tests on small Erdös-Renyi random graphs with $8$ nodes and $4$ free edges each. Among all, 5 tests (so 1.5\tiny E\footnotesize -4\%) have produced $4$ iterations - hitting the barrier of $f$ iterations. } \label{iterations}
\end{figure}

%Let us now give some precisions about how conjecture \ref{conj1} has been tested.
To test Conjecture \ref{conj1}, we have first generated random instances of increasing size of PRO and have recorded the number of iterations. Figure \ref{iterations} (left) shows that the number of iterations of PRI seems to grow at most linearly with the number of free edges. On the figure, random instances have been generated using a \textit{power-law} distribution \cite{aiello2001} but identical simulations have also been performed on \textit{Erd\"os-R\'enyi} random graphs \cite{erdos1960} or on portions of the real web, with about the same tendency each time.

In a second time, we have tried to generate instances that would perform more than $f$ iterations. Therefore, we have generated more than 200 million Erd\"os-R\'enyi and Power-law random instances, with parameters ranging from 3 to 10 free edges, 5 to 15 nodes and a highly variable number of edges, without ever being able to find such an example. Figure \ref{iterations} (right) shows how the number of iterations of PRI are distributed when generating many instances with $n = 8$ and $f = 4$. Note that we have also been exploring bigger value for $f$ and $n$ but since PRI behaves so well in practice, we have only been able to obtain a few iterations w.r.t. the problem size for these instances (always less than 8 iterations). By concentrating on small instances, we were able to generate some examples that were close to cross that $f$-iterations bound. Even if crossing this bound was possible, our simulations give a good indication about the scarcity of such examples when considering random graphs. Showing that random instances for which PRI takes more than $f$ iterations are unlikely to be observed is in our plans for further research.

%To test Conjecture \ref{conj1}, we have generated a significant number of random instances on which we have systematically verified its validity. In total, more than 200 million Erdös-Renyi and Power-law instances have been generated, with parameters ranging from 3 to 10 free edges, 5 to 15 nodes and a highly variable number of edges. We have also been exploring bigger value for $f$ and $n$ but since PRI behaves so well in practice, we have not been able to reach a significant enough number of iterations on such instances. By concentrating on small instances, we were able to generate some examples that were close to contradict Conjecture \ref{conj1}, even though they never did. Of course, testing every possible graph is not realistic and even a significant number of tests does not mean that counter-examples do not exist. However, if any, it could be a good indication about the scarcity of such examples, at least when considering random graphs. Figure \ref{iterations} shows how the number of iterations of PRI grows when increasing the problem size and how it is distributed when we run it on multiple random graphs of the same size.

\section{Particular cases}

In this section, we formulate some particular cases of PRO on which it can be shown that PRI behaves well. In many applications, it is assumed that the random walk used to compute PageRank can be interrupted at any time with some fixed probability $c$ and start again from an arbitrary node of the graph \cite{Berkhin2005}. This restarting probability is called \textit{zapping}. It can be seen as if the random surfer could get bored of performing its search with probability $c$ and decide to start a new search from a new randomly chosen node. We show below that in such cases, PRI converges in weakly polynomial time.

%In a first case, we consider PRO in which we add \textit{zapping}, meaning that at any time, the random walk can stop with some fixed probability $c$ and start again from an arbitrary node of the graph. This well-known variation of PageRank (see e.g. \cite{Berkhin2005}) can be interpreted as if the random websurfer could get bored of performing its search with probability $c$ and decide to start a new search from a new randomly chosen node. In that case, the following result holds.

\begin{thm}

PRO with fixed non-zero zapping probability $c$ can be solved in weakly polynomial time using PRI.

\begin{proof}
Our proof relies on results from Tseng and Puterman \cite{Tseng1990,Puterman1994}. Puterman shows that PI converges always in less iterations than the other well-known algorithm Value Iteration (VI). Furthermore, Tseng shows that VI converges in at most $O(n \log(n \delta) \, \eta^{-r})$, where $n$ is the number of states, $\delta$ is the binary input size, $\eta$ is the minimum non-zero transition probability and $r$ is the minimum number of steps needed to join two arbitrary nodes. In case of zapping, there is always a non-zero probability for any node to reach any other node in only one step, i.e. when zapping happens. Hence $r = 1$. Besides, because of the uniform random walk, $\eta$ is always at least $c/n$. Regrouping all the arguments, we show that PI must converge in at most $O(n^2 \log(n\delta) /c)$ steps, which is weakly\footnote{The bound is only \textit{weakly} polynomial because it depends on the number of bits that are necessary to represent the numbers in a problem instance.} polynomial in $n$ for a fixed value of $c$.
\vspace{-.6cm}

\end{proof}

\end{thm}

In the next case, we show that PRI converges in at most $f$ iterations when all free edges come out of the same arbitrary node $w$. Note that a particular case of this result was one of the main contribution of \cite{DeKerchove2008} : they were able to formulate an explicit optimal strategy when all edges come out of the starting node $v_s$.
%Note that a particularized version of that case has already interested de Kerchove et al. in \cite{DeKerchove2008} : they were able to formulate an explicit optimal strategy when all edges come out of the starting node $v_s$.
%For a subcase of the following one (in which all edges leave the target node), an optimal explicit strategy has already been found by de Kerchove et al. \cite{DeKerchove2008}. We show that in that case, PRI also finds this optimal strategy quickly and we generalize the situation by allowing edges to leave either the target node, or one other arbitrary node.

\begin{thm}

PRI takes less than $f$ iterations when all the free edges go out of the same node $w$ and/or out of the starting node $v_s$.

\begin{proof}

We are going to show that in all considered cases, PRI always makes at least one final decision in each step (final in the sense that it will never be undone in a subsequent iteration). If this is true, then of course PRI takes at most $f$ iterations since we may consider at least one less free edge at each iteration. Furthermore, observe that the nodes may always be sorted w.r.t. their first hitting time at each iteration of PRI~: this will be the key to derive our result. The proof goes in three steps~: first we suppose that all free edges leave node $v_s$, then that they all leave some other node $w$ and we finally unify these two results to prove the claim.
\begin{itemize}

	\item \textit{\textbf{Case 1~:} all free edges leave node $v_s$.} Since no edge enters $v_s$, switching a free edge that leaves $v_s$ does not influence the first hitting times of the other nodes (it does not shorten or lengthen their path). Hence, their first hitting times is fixed from the beginning and only $\p_{v_s}$ decreases in the iterative process. Suppose that every free edge is initially activated. Then, since $\p_{v_s}$ can only decrease, $\p_{v_s} - (\p_u + 1)$ can also only decrease for any node $u$ such that $(v_s, u) \in \F$, and therefore free edges can only be deactivated by PRI (see line 3 of the algorithm). So PRI never undoes any of its choices and it converges in at most $f$ iterations. If the initial policy was different, the argument is the same except at the first step where free edges $(v_s, u)$ such that $\p_{v_s} \geq (\p_u + 1)$ are activated and the other free edges are deactivated. Then again, free edges can only be deactivated since $\p_{v_s}$ is the only one to decrease.
	
	\item \textit{\textbf{Case 2~:} all free edges leave some node $w \neq v_s$.} Here, the key is to see that when switching a free edge, $\p_w$ decreases more than the other nodes' first hitting times. If this is true, then it means that for any node $u \neq w$, $\p_w - (\p_u + 1)$ can only decrease and that free edges can only be deactivated at each step, so the argument used in case $1$ is still valid. Hence PRI would again take at most $f$ iterations. It only remains to prove that $\p_w$ indeed decrease faster than any other first hitting time, which we do next.
	
	Let us consider any node $u \neq w$. Among all the paths starting from $u$ that lead to the target node $v_t$, only those that go through $w$ will be shortened when switching free edges, since all free edges leave $w$. Let us thus partition the set of all paths from $u$ to $v_t$ into the ones that go through $w$ that we denote by $P_{uwv}$, and the ones that do not go through $w$ that we denote by $P_{uv}$. We also denote the average weighted length of the paths in $P_{uv}$ by $\p_{uv}$ and the probability to take such a path by $p_{uv}$, all w.r.t the probability for the considered paths to be chosen. If a path goes through $w$, it means that $u$ reaches $w$ before reaching $v_t$ (since $v_t$ is absorbing). Therefore, the probability of hitting $w$ before hitting $v_t$ is given by $p_{uw} = 1 - p_{uv}$, and we denote the average weighted length of paths between $u$ and the first visit of $w$ by $\p_{uw}$. Using these notations, we can write the first hitting time of $u$ as follows :
	\begin{equation} \label{eq1}
		\p_u^{\mu_k} = p_{uv} \p_{uv} + (1 - p_{uv}) (\p_{uw} + \p_w^{\mu_k})
	\end{equation}
	where $(\p_{uw} + \p_w^{\mu_k})$ is the average weighted length of a path that goes through node $w$ before reaching $v_t$. In this equation, observe that only $\p_w^{\mu_k}$ can change during the iterative process of PRI since the changes to the probability distributions and to the average lengths of paths can only happen when travelling through $w$. Let us now suppose that in some step $k$ of PRI, $\p_w^{\mu_k}$ decreases from $\Delta \p^k$, so $\p_w^{\mu_{k+1}} = \p_w^{\mu_k} - \Delta \p^k$. Using equation (\ref{eq1}), the influence of this decrease on $\p_u^{\mu_k}$ is thus :
	\begin{equation*}
		\p_u^{\mu_{k+1}} = \p_u^{\mu_k} - (1 - p_{uv}) \Delta \p^k.
	\end{equation*}
	Hence, $\p_u^{\mu_k}$ decreases of at most $\Delta \p^k$, but only if all its paths to $v_t$ pass through $w$. Therefore, the first hitting time of $w$ decreases more than the first hitting time of any other node. This concludes the proof for this case.
%	But then, $\p_w - (\p_u + 1)$ can only decrease and free edges can only be deactivated at each step, so the argument used in case $1$ is still valid. Hence PRI again takes at most $f$ iterations.
	
	\item \textit{\textbf{Case 3~:} all free edges leave either $v_s$ or $w$.} Since node $w$ is not influenced by node $v_s$, we can consider the PRI process in $w$ independently from the process in $v_s$. Thus, in node $w$, applying case 2, PRI makes at least one switch that is final in each step until every free edge leaving $w$ reaches its optimal activation state. At that point, the first hitting times of every node is fixed for the rest of the process except maybe in node $v_s$. If $\p_{v_s}$ has not reached its optimal value yet, we let PRI run as if we were in case 1. So, we first focus on $w$ and observe that at least one final switch is made there at each step until the optimal configuration of the free edges leaving $w$ is reached. And then we focus on node $v_s$ where the same observation can be made. Combining both subprocesses, we conclude that one final decision is made at each iteration and so, again, PRI takes at most $f$ iterations to converge.
	\vspace{-.9cm}
	
\end{itemize}

\end{proof}

\end{thm}

\bibliographystyle{alpha}
\bibliography{biblio}

\end{document}